\documentclass[11pt]{article}
\usepackage[margin=1in]{geometry}
\usepackage{amsmath,amssymb}
\usepackage{graphicx}
\usepackage{booktabs}
\usepackage{array}
\usepackage[version=4]{mhchem}
\usepackage[detect-all]{siunitx}
\usepackage[numbers,sort&compress]{natbib}
\usepackage{microtype}
\usepackage{xcolor}
\usepackage[colorlinks=true,linkcolor=blue!50!black,citecolor=blue!50!black,urlcolor=blue!50!black]{hyperref}
\newcommand{\cho}{\ce{CH3O2}}
\title{Kinetic versus refill-zone uncertainty in predicted pre-burning of methane–oxygen rotating detonation rocket engines}
\author{Y. Sungtaek Ju\thanks{University of California, Los Angeles, CA 90095-1597; sungtaek.ju@ucla.edu}}
\date{}

\begin{document}
\maketitle

\begin{abstract}
Rotating detonation rocket engines can lose performance when fresh propellant mixes with hot recirculated combustion products and ignites before the detonation wave arrives. Predicting this pre-burning requires fuel ignition chemistry at high pressure and in product-diluted mixtures, conditions most chemical kinetic mechanisms used in engine simulations were not developed for. With methane as our base fuel, we quantify how strongly predicted pre-burning depends on the choice of mechanism, test the mechanisms against published high-pressure ignition data, and use Bayesian methods to calibrate the most influential uncertain reaction and to assess which measurements would best improve the predictions. As the refill zone is represented by an idealized zero-dimensional model in the present study, our results should be viewed as relative comparisons rather than direct engine predictions. Five existing chemical kinetics mechanisms for methane combustion predict refill-zone ignition delays that differ by up to a factor of twenty. Mechanisms without methylperoxy (\cho{}) chemistry, including two commonly used in engine simulations, do not reproduce the laboratory data below about 1150~K. Calibrating the \ce{CH3 + HO2} reaction against these data, while allowing for systematic differences between datasets, increases the relative importance of its chain-terminating channel, consistent with a direct rate measurement. With the data-consistent mechanisms, the remaining kinetic uncertainty in the predicted pre-burnt fraction is, in most cases, several times smaller than the uncertainty arising from the state of the refill zone: the temperature and composition of the recirculated products and the way fresh and burnt gases mix. This comparison does not include alternative carbon chemistry under radical seeding, which the data do not test directly. Further kinetic experiments of the types considered would reduce its parametric part only slightly. Improved pre-burning predictions are therefore expected to depend mainly on accurately characterizing the refill zone.
\end{abstract}

\section{Introduction}
Rotating detonation engines (RDEs) are promising for compact, lightweight rocket propulsion. Additively manufactured rotating detonation thrust chambers have been hot-fire tested to thermal steady state in burns lasting up to 133~s \cite{Teasley2023}. In the refill zone, fresh propellant contacts hot products recirculated from the previous cycle, and part of it is thought to burn before the detonation wave arrives. Reduced-order loss models identify this pre-detonation deflagration as the leading loss term \cite{Huff2025}.

Previous analyses have estimated pre-burning from ignition-delay calculations. Stechmann et al.\ \cite{Stechmann2019} compared homogeneous-reactor ignition delays with cycle times and argued that pre-ignition between waves grows with chamber pressure; using GRI-Mech~3.0, they expected it to become important for \ce{CH4}/\ce{O2} at pressures of the order of 100~atm. Fievisohn et al.\ \cite{Fievisohn2019} modeled the constant-pressure mixing of recirculated products into the reactants. They showed that the thermodynamic benefit of detonation decreases with the product fraction, but also noted that the detailed mechanisms they used are not validated at these conditions. In a time-dependent reactor model with expansion, mixing and chemistry, they later found that autoignition at higher product fractions shifts the wave from a strong mode (85--100\,\% of the Chapman--Jouguet speed) to a marginal mode (60--70\,\%), consistent with wave speeds measured in a premixed hydrogen--air RDE \cite{Fievisohn2020}.

These estimates depend on the ignition delay of fresh \ce{CH4}/\ce{O2} diluted with hot, \ce{H2O}- and \ce{CO2}-rich products at \SIrange{10}{100}{atm} and about \SIrange{1000}{1400}{K}. These conditions lie outside the validation range of the mechanisms commonly used in RDE simulations, including GRI-Mech~3.0 \cite{GRI30}, FFCM-1 \cite{FFCM1} and a recent 10-step global mechanism \cite{Yungster2026}. Work on high-pressure methane ignition, including for supercritical-\ce{CO2} power cycles, has shown that methylperoxy (\cho{}) chemistry is important for methane ignition at high pressure \cite{Karimi2019,HarmanThomas2022,HarmanThomas2023,Hashemi2016,Petersen1999CF}. Key \cho{} rate coefficients, however, remain relatively poorly known at combustion temperatures \cite{HarmanThomas2023,Tsang1986}. Each of the engine analyses above relies on a single kinetic mechanism, and the sensitivity of the results to this choice of mechanism has not been assessed.

This paper examines how much the chemistry contributes to the uncertainty in predicted pre-burning, and which measurements would reduce that uncertainty most. First, we compare five established mechanisms and evaluate the predicted pre-burnt fraction and ideal specific-impulse gain. Second, we test the mechanisms against existing high-pressure ignition data and calibrate the most influential rate constants with a hierarchical Bayesian model, which allows for systematic differences between datasets and laboratories. We compare the result with a direct rate measurement and theoretical prediction. Third, goal-oriented Bayesian experimental design \cite{HuanMarzouk2013,Huan2024,WangSheen2015} is employed to rank candidate experiments by how much they would reduce the uncertainty in the pre-burnt fraction.  We then compare this kinetic uncertainty with that arising from the state of the refill zone.

The engine side of the present analysis is idealized. The refill zone is zero-dimensional: fresh \ce{CH4}/\ce{O2} ($\phi = 1.1$, \SI{300}{K}) mixes adiabatically at constant pressure with equilibrium products at \SI{3000}{K}, and two alternative product states serve as a sensitivity test. The distribution of mixing states is prescribed by two phenomenological models rather than directly simulated. Wave interaction, heat loss and other equivalence ratios are not considered. The pre-burnt fraction $\beta$ is therefore considered a model metric rather than a measurable quantity. The specific-impulse gain we calculate is an ideal upper bound. On the chemistry side, the laboratory data are simulated as ideal homogeneous reactors, and the calibration varies only the two channels of \ce{CH3 + HO2}; the calibrated rate factors should therefore be regarded as effective values that may compensate for other deficiencies. With these approximations, our results provide relative comparisons of kinetic and refill-zone uncertainty within this framework, not predictions for a specific engine.

\section{Methods}
\subsection{Kinetic mechanisms}
We compare five mechanisms for methane combustion: GRI-Mech~3.0 \cite{GRI30} (53 species, 325 reactions); FFCM-1 \cite{FFCM1} (38 species, 291 reactions); FFCM-2 \cite{FFCM2,Zhang2024} (96 species, 1054 reactions), which comes with a 258-dimensional posterior covariance of its rate parameters; the University of Sheffield mechanism UoS~sCO$_2$~2.1 \cite{HarmanThomas2023,HarmanThomasThesis} (114 species, 817 reactions), built from the published CHEMKIN files after correcting formatting errors only; and the NASA 10-step global mechanism \cite{Yungster2026} in its version optimized for 10~bar. GRI-Mech~3.0, FFCM-1 and the 10-step mechanism contain no \cho{}, and the 10-step mechanism contains neither \ce{CH3} nor \ce{HO2}. As a diagnostic, we also use FFCM-2 with its 33 \cho{} reactions removed. Our implementation of the 10-step mechanism reproduces the published ignition delays to within 5\,\% at 1400--1800~K. To compare it with diluted laboratory data, we added Ar, \ce{N2} and He to it as inert species without changing any rate. All calculations use Cantera~3.2 \cite{Cantera} with ideal-gas thermodynamics.

\subsection{Refill-zone ignition}
Fresh \ce{CH4}/\ce{O2} ($\phi = 1.1$, $T_f = \SI{300}{K}$ unless stated otherwise) is mixed at constant pressure with a mass fraction $\psi$ of equilibrium products ($\phi = 1.1$, \SI{3000}{K}). The resulting mixture temperatures range from \SI{680}{K} at $\psi = 0.1$ to \SI{1770}{K} at $\psi = 0.5$. The ignition delay $\tau(\psi)$ is defined as the time, at constant pressure, to reach half of the temperature rise to equilibrium. A criterion based on the maximum rate of temperature rise is unsuitable here: in these radical-bearing mixtures, radical recombination raises the temperature by 50--170~K within about \SI{1}{\micro\second}, and such a criterion can mistake this transient for ignition. Because recirculated products have expanded and partly recombined before they mix, two alternative product states bracket the baseline: equilibrium at \SI{2500}{K} (cooler, with fewer radicals), and the \SI{3000}{K} equilibrium composition with the radicals H, O, OH, \ce{HO2} and \ce{H2O2} removed (same temperature, no radical seeding).

\subsection{Pre-burnt fraction and ideal cycle}
Fresh propellant enters uniformly over the fill time $t_\mathrm{fill}$ (\SIlist{50;200}{\micro\second}). A parcel pre-burns if it ignites before the wave arrives, and the pre-burnt fraction $\beta$ is the mass fraction of fresh propellant that pre-burns. Two mixing models describe how parcels acquire products. In model~A, parcels are mixed instantly, with a product fraction $\psi$ drawn from a beta distribution of mean $\bar\psi$. In model~B, used for the main results, each parcel entrains products progressively, $\psi(t) = 1 - e^{-\omega t}$, with a lognormally distributed rate $\omega$ chosen to give the mean product fraction $\bar\psi$ at wave arrival. Ignition in model~B is evaluated with the Livengood--Wu integral \cite{LivengoodWu1955}. To test this closure, we also integrate entrainment and chemistry together for single parcels on a grid of $\omega$ (operator splitting; ignition when the temperature exceeds the frozen mixing temperature by half of the rise to equilibrium, consistent with the definition of $\tau$) and evaluate $\beta$ with these ignition times.

The ideal specific-impulse gain of detonation over constant-pressure combustion is computed for the Chapman--Jouguet state of each parcel, followed by isentropic expansion in chemical equilibrium ($p_1/p_e = 1000$). Pre-burnt parcels are assumed to contribute no specific-impulse gain.

\subsection{Experimental data}
Table~\ref{tab:data} lists the ignition-delay data used in the present study. Shock-tube data are simulated at constant volume with the ignition definition of each record: onset by extrapolation of the steepest rise, time of maximum slope, or, for the fuel-rich data of \cite{Shao2020}, 50\,\% methane consumption. Data defined by a threshold on a relative species signal are excluded (102 points), as are points with species missing from a mechanism. The rapid-compression-machine (RCM) data \cite{Hashemi2016} are simulated as adiabatic constant-volume reactors starting from the compressed state, because the effective-volume histories are not published. This omits heat loss and biases every mechanism toward short delays.

\begin{table}[ht]\centering\small
\caption{Ignition-delay data. Groups G1--G4 are the shock-tube groups used in the calibration. Column `calib.' gives the observations used there: G1 at $p \geq 10$~atm and $T_5 \leq 1500$~K, and all of G2--G4, excluding seven that do not ignite in the simulation. The RCM set is used only for a one-sided test.}\label{tab:data}
\resizebox{\textwidth}{!}{%
\begin{tabular}{llrrll}
\toprule
group & source & audit & calib. & conditions & definition \\ \midrule
G1 & ChemKED methane shock tube \cite{ChemKED} & 775 & 311 & up to 264~atm, Ar/\ce{N2}/He, some \ce{CO2}/\ce{H2O} & per record \\
G2 & Shao~2019 \cite{Shao2019}, Barak~2020 \cite{Barak2020} & 27 & 27 & 27--286~atm, 36--86\,\% \ce{CO2} & onset \\
G3 & Karimi~2019 \cite{Karimi2019}, Pryor~2017 \cite{Pryor2017JERT} & 40 & 40 & 7--209~atm, \ce{CO2} and Ar & max.\ slope / CH* peak \\
G4 & Shao~2020 \cite{Shao2020} (not an FFCM-2 target) & 24 & 24 & 9--53~atm, 1440--1855~K, Ar & onset / 50\,\% \ce{CH4} \\
RCM & Hashemi~2016 \cite{Hashemi2016} & 76 & -- & 15--80~bar, 900--1200~K & pressure \\
\bottomrule
\end{tabular}}
\end{table}

\subsection{Calibration of the \ce{CH3 + HO2} channels}
\paragraph{Screening.} Rate variants of FFCM-2 are created by multiplying reaction 108 (\ce{CH3 + HO2 -> CH4 + O2}, the termination channel), reaction 109 (\ce{CH3 + HO2 -> CH3O + OH}, the propagation channel; numbering of the FFCM-2 mechanism file) or the \cho{} reactions by fixed factors. For a first, deliberately conservative screening, each data file contributes the mean Gaussian log-likelihood of its points, with residual $r = \ln(\tau_\mathrm{exp}/\tau_\mathrm{model})$, the stated uncertainty $s$, and a model discrepancy $\sigma_d = 0.36$.

\paragraph{Hierarchical calibration.} The parameters are the logarithms of the two rate factors, $\boldsymbol x = (\ln m_\mathrm{term}, \ln m_\mathrm{prop})$. Each of the 402 calibration observations (Table~\ref{tab:data}) was simulated on a $3\times3$ grid of factors between $1/3$ and 3, and a quadratic surrogate of $\ln\tau$ was fitted (validation error at most 0.05 in $\ln\tau$). For observation $i$ of dataset $j$ (90 datasets), the data model is
\begin{equation}
y_i = f_i(\boldsymbol x) + b_j + c_j\,\bigl(1000/T_i - \overline{1000/T}_j\bigr) + \varepsilon_i,
\end{equation}
with $b_j \sim \mathcal N(0,\tau_b^2)$, $c_j \sim \mathcal N(0,\tau_c^2)$ and $\varepsilon_i \sim \mathcal N(0, s_i^2 + \sigma_{w,g}^2 + u_i^2)$. Here, $y_i$ is the measured $\ln\tau$, $b_j$ is a dataset offset (apparatus, calibration and ignition-definition effects), $c_j$ a dataset drift in inverse temperature (for example from a pressure rise behind the reflected shock at long delays), $\sigma_{w,g}$ additional scatter within data group $g$, and $u_i$ the surrogate error. The random effects are integrated out analytically, and the remaining parameters are sampled by adaptive Metropolis with $\boldsymbol x \sim \mathcal N(0,1)$ truncated to the surrogate domain and half-normal(0.5) priors on the scales. We refer to this model (model~D in Appendix~\ref{app:calib}) as the \emph{dataset-drift model}. Simpler error models, five robustness variants (including a laboratory-level offset and FFCM-2's own posterior as prior), their comparison by leave-one-out cross-validation, and the convergence diagnostics are described in Appendix~\ref{app:calib}. The \cho{} rates are held at their FFCM-2 values; within their literature uncertainty they change $\beta$ by at most 0.6 percentage points (Section~\ref{sec:branching}). The direct measurement of \cite{Hong2012} is added as data on both factors by importance re-weighting of the posterior samples. Because Hong et al.\ do not state a confidence level for their uncertainty factors (1.4 and 2.1), we read them as one, two and three standard deviations in turn \cite{Nagy2015}. The same procedure is applied to the two \ce{CH3 + HO2} channels of UoS~sCO$_2$~2.1, and the two mechanisms are compared on the 379 observations that both can simulate.

\subsection{Experimental design}
Candidate experiments are (i) 138 feasible ignition-delay measurements of \ce{CH4}/\ce{O2} diluted with Ar, \ce{CO2} or radical-free product species at \SIrange{950}{1300}{K} and \SIrange{10}{100}{atm}, and (ii) nine rate experiments of the type of \cite{Hong2012}, in which \ce{OH} and an \ce{HO2}-sensitive absorbance are measured in \ce{H2O2}/\ce{CH4}/Ar mixtures at 1050--1250~K and 3.5--30~atm. The value of an experiment, or of a series of experiments in one apparatus, is the expected reduction of the variance of $\beta$, summed over 18 conditions (10, 30 and 100~atm; $\bar\psi = 0.1$, 0.2 and 0.3; fill times of 50 and \SI{200}{\micro\second}). It is estimated by the Monte Carlo method: outcomes are simulated from the calibrated posterior with a plausible noise model, and the posterior is re-weighted by each outcome. For rate experiments, the noise model includes the uncertainty of the interfering reactions, which dominated the error budget of \cite{Hong2012}, the uncertainties of temperature and initial \ce{H2O2} loading, and detection limits; it is detailed in Appendix~\ref{app:design}.

\subsection{Reporting of uncertainty}\label{sec:reporting}
We report each source of uncertainty as the change it causes in the predicted pre-burnt fraction, in percentage points. For uncertain parameters, the change is the range covering 90\,\% of the calibrated predictions; for alternative mechanisms or modeling assumptions, it is the difference between the highest and lowest prediction. Unless otherwise noted, the remainder of the paper quotes values at a reference condition (100~atm, $\bar\psi = 0.2$ and $t_\mathrm{fill} = \SI{200}{\micro\second}$), where the calibrated FFCM-2 with mixing model~B predicts $\beta \approx 25\,\%$. Values at all 18 conditions are given in Appendix~\ref{app:budget}.

\section{Results}
\subsection{The mechanisms disagree on refill-zone ignition}
\begin{figure}[ht]\centering
\includegraphics[width=\textwidth]{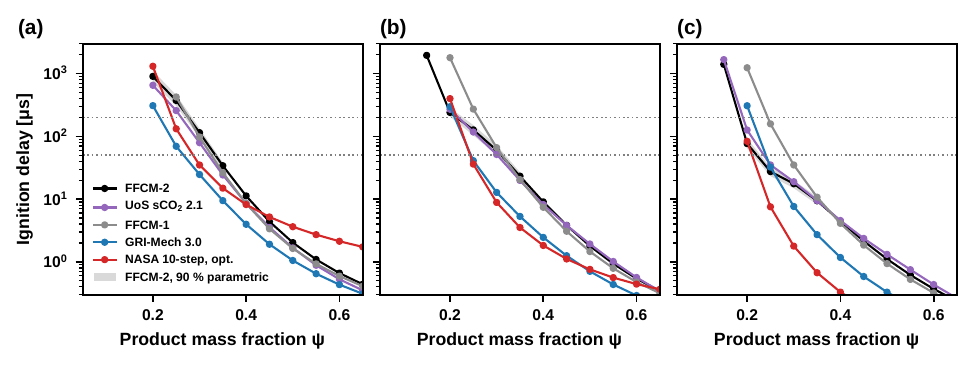}
\caption{Ignition delay of refill-zone mixtures (fresh \ce{CH4}/\ce{O2}, $\phi = 1.1$, \SI{300}{K}, plus a mass fraction $\psi$ of \SI{3000}{K} equilibrium products) at (a) 10, (b) 30 and (c) 100~atm for the five mechanisms. Gray band: 90\,\% uncertainty range of FFCM-2 from its parameter covariance; dotted lines: fill times of 50 and \SI{200}{\micro\second}.}\label{fig:tau}
\end{figure}

The five mechanisms predict refill-zone ignition delays that differ by up to a factor of about 20 (Fig.~\ref{fig:tau}). At 100~atm and $\psi = 0.25$, for example, the predicted delays are \SI{7.5}{\micro\second} (10-step), \SI{27.5}{\micro\second} (FFCM-2), \SI{32}{\micro\second} (GRI-Mech~3.0) and \SI{159}{\micro\second} (FFCM-1). The mechanisms also differ in how ignition scales with pressure. In FFCM-2, the product fraction required for ignition within \SI{50}{\micro\second} falls from 0.33 at 10~atm to 0.22 at 100~atm, so that pre-burning increases with pressure; in GRI-Mech~3.0 and FFCM-1 this threshold changes little.

The difference between FFCM-2 and FFCM-1 is due to methylperoxy chemistry: FFCM-2 without its \cho{} reactions reproduces FFCM-1 (\SI{174}{\micro\second} versus \SI{159}{\micro\second} at 100~atm and $\psi = 0.25$). By contrast, the parametric uncertainty of FFCM-2 itself is small (standard deviation of 0.08--0.13 in $\ln\tau$). At 100~atm and $\psi = 0.25$, FFCM-1 lies a factor of 5 beyond the upper end of the FFCM-2 uncertainty band. Before the laboratory data are considered, the choice of mechanism therefore dominates the kinetic uncertainty. Within FFCM-2, the termination channel \ce{CH3 + HO2 -> CH4 + O2} accounts for 41--49\,\% of the variance of $\ln\tau$.

\subsection{Consequences for the pre-burnt fraction and the specific impulse}
The predicted pre-burnt fraction ranges from about 3\,\% to 47\,\% and is set mainly by the mean product fraction $\bar\psi$. At the reference condition, raising $\bar\psi$ from 0.1 to 0.3 increases $\beta$ from 8\,\% to 41\,\%. The five mechanisms predict $\beta$ between 20\,\% (FFCM-1) and 29\,\% (10-step) at the reference condition (Fig.~\ref{fig:ensemble}a). The differences of up to a factor of 20 in ignition delay thus shrink to factors of at most 2.2 in $\beta$, because the distribution of mixing states smooths the ignition threshold. Section~\ref{sec:budget} compares these and other sources of uncertainty after the mechanisms have been tested against the laboratory data.

\begin{figure}[ht]\centering
\includegraphics[width=\textwidth]{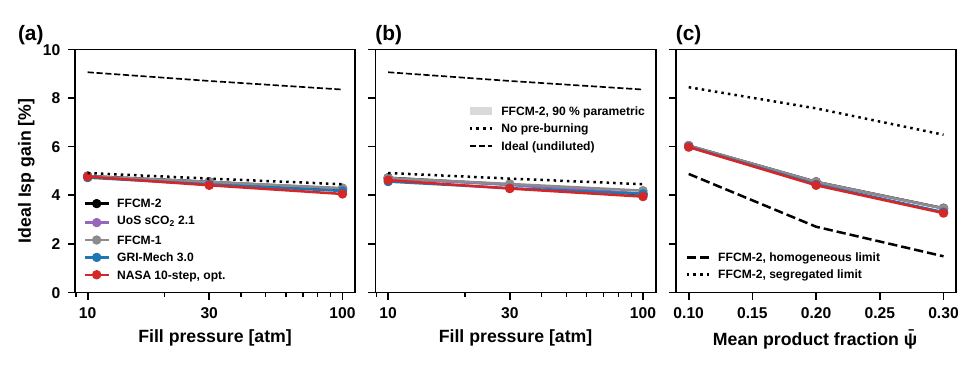}
\caption{Ideal specific-impulse gain of detonation over constant-pressure combustion with refill dilution and pre-burning (\ce{CH4}/\ce{O2}, $\phi = 1.1$, $p_1/p_e = 1000$). (a, b) Parcel-resolved gain versus fill pressure at $\bar\psi = 0.2$ for fill times of (a) 50 and (b) \SI{200}{\micro\second}: five mechanisms, 90\,\% uncertainty range of FFCM-2, the gain without pre-burning (dilution only), and the ideal undiluted gain; the five mechanism curves nearly coincide. (c) Gain versus mean product fraction at 30~atm and \SI{50}{\micro\second}, with the FFCM-2 homogeneous and segregated mixing limits.}\label{fig:cycle}
\end{figure}

The ideal specific-impulse gain of detonation over constant-pressure combustion is 9.1, 8.7 and 8.4\,\% at 10, 30 and 100~atm. Dilution of the fresh charge by recirculated products roughly halves it (from 8.7\,\% to 4.7\,\% at 30~atm, $\bar\psi = 0.2$ and \SI{50}{\micro\second}), consistent with the trend found by Fievisohn et al.\ \cite{Fievisohn2019}. Pre-burning reduces the gain by at most a further 0.6 percentage points, because the parcels that pre-burn are the most diluted ones and would have contributed little gain. The choice of mechanism changes the gain by at most 0.3 percentage points (Fig.~\ref{fig:cycle}).

\paragraph{Why the pre-burnt fraction matters beyond the ideal impulse.} The rather impulse cost of pre-burning in the parcel model results from the assumption that each parcel keeps its own composition, so that pre-burning removes mainly the most diluted parcels. If pre-burnt pockets remain segregated from the charge and contribute no gain, each percentage point of $\beta$ costs up to 0.09 percentage points of ideal gain; if they mix into the charge ahead of the wave, it costs up to 0.17 (Fig.~\ref{fig:cycle}c). In the latter case a reacted fraction of only 5\,\% halves the Chapman--Jouguet pressure ratio at 30~atm. Even then, the differences between mechanisms correspond to at most about one percentage point of ideal gain, so the kinetic uncertainty of $\beta$ matters little for the ideal cycle. Its importance lies in effects outside the ideal cycle. Autoignition ahead of the wave can change the wave itself, from a strong to a marginal mode \cite{Fievisohn2020}, and reduced-order loss models attribute 40--80\,\% of the detonation losses to parasitic deflagration, with the largest effect on the pressure ratio across the wave \cite{Huff2025}. Neither effect is represented in our cycle estimate. Further details are given in Appendix~\ref{app:budget}.

\subsection{Comparison with experimental data}\label{sec:audit}
\begin{figure}[ht]\centering
\includegraphics[width=\textwidth]{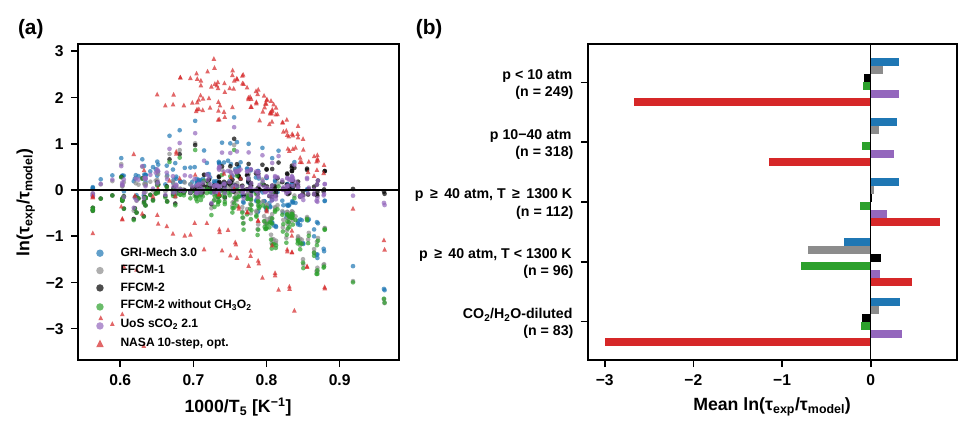}
\caption{Comparison with the ChemKED methane shock-tube data. (a) Residuals $\ln(\tau_\mathrm{exp}/\tau_\mathrm{model})$ at $p \geq 40$~atm versus $1000/T_5$; triangles: NASA 10-step mechanism (points where it ignites). (b) Mean residual by region (colors as in (a); $n$ counts the points simulated by all detailed mechanisms, and the 10-step means use those of them where it ignites); positive values: the mechanism ignites too fast.}\label{fig:audit}
\end{figure}

\paragraph{Overall agreement.} Table~\ref{tab:audit} and Fig.~\ref{fig:audit} summarize how well each mechanism reproduces the laboratory data. Against the ChemKED shock-tube data (G1, 775 points), FFCM-2 agrees best, followed by UoS~sCO$_2$~2.1, FFCM-2 without \cho{}, FFCM-1, and GRI-Mech~3.0. The differences are largest where \cho{} chemistry matters, at $\geq 40$~atm and below \SI{1300}{K}. There, the two mechanisms with \cho{} reproduce the measured delays to within about 10\,\% on average, whereas FFCM-1 and FFCM-2 without \cho{} ignite about twice as slowly and GRI-Mech~3.0 about 1.35 times as slowly. In the \ce{CO2}-diluted data above 200~atm (G2), the mechanisms without \cho{} ignite 2 to 4.6 times more too slow. The NASA 10-step mechanism deviates much more than the detailed mechanisms and does not ignite
within 20 measured delays at 241 of the 775 points.

\begin{table}[ht]\centering\small
\caption{Agreement with the ignition data. Columns 3--5 give the mean ratio of simulated to measured ignition delay (geometric mean; values above 1: the mechanism ignites too slowly). The RCM comparison omits heat loss, which would lengthen every simulated delay.}\label{tab:audit}
\begin{tabular}{lcccc}
\toprule
 & G1, all & G1, $\geq 40$~atm, & G2, & RCM, $\geq 38$~atm, \\
mechanism & rms of $\ln(\tau_\mathrm{exp}/\tau_\mathrm{model})$ & $<$\SI{1300}{K} & $>200$~atm & delay $<$\SI{20}{ms} \\ \midrule
FFCM-2 & 0.40 & 0.90 & 0.63 & 0.65 \\
UoS sCO$_2$ 2.1 & 0.48 & 0.90 & 0.93 & 0.86 \\
FFCM-1 & 0.58 & 2.0 & 4.4 & 3.9 \\
FFCM-2 without \cho{} & 0.57 & 2.2 & 4.6 & 3.9 \\
GRI-Mech 3.0 & 0.59 & 1.35 & 2.0 & 2.5 \\
NASA 10-step & 1.9$^a$ & 0.63$^a$ & 8.8 & -- \\
\bottomrule
\end{tabular}\\[2pt]
{\footnotesize $^a$Points where the mechanism ignites.}
\end{table}

\paragraph{Evidence for methylperoxy chemistry and its limits.} The evidence that \cho{} chemistry is needed rests on a small number of independent sources. Most of the shock-tube points that discriminate between FFCM-2 and its \cho{}-free variant come from Petersen et al.\ \cite{Petersen1999JPP}, whose data were also used to develop FFCM-2, and from Huang et al.\ \cite{Huang2004} at 1000--1200~K. Near 1200--1300~K and 25--48~atm, the data of \cite{Burke2015,Mathieu2015} point the other way: there FFCM-2 ignites up to 1.8 times too fast, and the \cho{}-free variant agrees better. On the shock-tube data not used to develop FFCM-2, FFCM-2 still agrees best overall, but the few such points in the \cho{}-sensitive region favor the \cho{}-free mechanisms. Long delays at low temperature are also where a gradual pressure rise in the shock tube shortens the measured delay and mimics extra reactivity. Re-simulating the 311 G1 calibration points with a pressure rise of 2--4\,\% per ms leaves the evidence essentially unchanged: the \cho{}-free variant remains about twice too slow in the \cho{}-sensitive region. Appendix~\ref{app:audit} gives the point counts and the pressure-rise results.

\paragraph{The data-consistent mechanisms are complementary.} UoS~sCO$_2$~2.1, which was tuned to supercritical-\ce{CO2} data, agrees best with the \ce{CO2}-diluted data of G2, where FFCM-2 ignites 1.6 times too fast above 200~atm. FFCM-2 agrees best with the argon and nitrogen data, including the high-temperature set G4, which was not used in its development (rms 0.14 versus 0.30). Each mechanism thus agrees best with the kind of data it was tuned to. This is the expected outcome of parameter tuning and does not by itself show that the differences between the mechanisms are structural.

\paragraph{One-sided test with RCM data.} In the RCM data, the residual of every mechanism grows with delay time, by about a factor of three from delays below \SI{5}{ms} to delays above \SI{20}{ms}, as expected from the neglected heat loss. Because heat loss can only lengthen simulated delays, the comparison provides a one-sided test. At $\geq 38$~atm and delays below \SI{20}{ms}, the mechanisms without \cho{} are already 2.5--3.9 times too slow before any correction, whereas those with \cho{} are too fast (Table~\ref{tab:audit}). Starting the simulations from the compressed state also omits chemistry during compression, which biases simulated delays in the same direction as the \cho{} deficit. This effect is expected to be small at these temperatures compared with the observed factors of 2.5--3.9, but it means the test is not strictly one-sided. Within this caveat, the RCM data are inconsistent with \cho{}-free chemistry; they cannot rank the remaining mechanisms.

\paragraph{The \cho{} effect does not require radical seeding.} In argon-diluted mixtures without radicals, removing \cho{} lengthens ignition by a factor of 4.3--4.6 at 30~atm and \SI{1050}{K}, and by 8.2--9.2 at 100~atm and \SI{1000}{K}, at delays of 1--5~ms. Replacing argon by \ce{H2O} or \ce{CO2} changes these ratios little, consistent with \cite{HarmanThomas2023}.

\paragraph{Reliability of the FFCM-2 uncertainty band.} The uncertainty band of FFCM-2 does not represent its systematic error in this regime. At $\geq 40$~atm and below \SI{1300}{K}, FFCM-2 ignites faster than all 13 of its own methane ignition targets, by factors of 1.08--1.69. Eleven of these deviations exceed the uncertainty reported by the developers (factors of 1.08--1.10). The deviations lie mostly within the experimental uncertainties, and the 13 targets come from only two studies \cite{Petersen1999JPP,Shao2019}. The evidence is therefore the common sign of the deviations rather than their size.

\subsection{The remaining kinetic uncertainty: the \ce{CH3 + HO2} branching}\label{sec:branching} Two candidates for the remaining kinetic uncertainty are the \cho{} subsystem and the \ce{CH3 + HO2} reaction. Perturbing the \cho{} reactions of FFCM-2 within their literature uncertainty (factors of 3, and 10 for \ce{CH3O2 + CH4}) changes $\beta$ by at most 0.6 percentage points. The rate of \ce{CH3O2 + CH4} is not a key uncertainty here: a recent theoretical value \cite{Xu2022} is 0.7--2.2 times the values in FFCM-2 and UoS~sCO$_2$~2.1 at 1000--1300~K. The two channels of \ce{CH3 + HO2}, by contrast, matter: a faster termination or a slower propagation channel lowers $\beta$ by up to 2--4 percentage points.

Table~\ref{tab:rates} compares the two channels in the mechanisms, in theory \cite{Jasper2009} and in the shock-tube measurement of Hong et al.\ \cite{Hong2012}. Both mechanisms lie close to the theory. Relative to FFCM-2 at \SI{1150}{K}, the measurement corresponds to a termination channel twice as fast and a propagation channel 0.72 times as fast, so it favors termination considerably more than the mechanisms do.

\begin{table}[ht]\centering\small
\caption{\ce{CH3 + HO2} rate coefficients at \SI{1150}{K} (\si{cm^3.mol^{-1}.s^{-1}}).}\label{tab:rates}
\resizebox{\textwidth}{!}{%
\begin{tabular}{lccc}
\toprule
 & \ce{-> CH3O + OH} (propagation) & \ce{-> CH4 + O2} (termination) & ratio \\ \midrule
FFCM-2 & \num{9.4e12} & \num{2.2e12} & 4.3 \\
UoS sCO$_2$ 2.1 & \num{9.9e12} & \num{2.9e12} & 3.4 \\
Jasper et al.\ 2009 (theory) \cite{Jasper2009} & \num{9.0e12} & \num{2.9e12} & 3.1 \\
Hong et al.\ 2012 (1054--1249 K) \cite{Hong2012} & \num{6.8e12} (factor 1.4) & \num{4.4e12} (factor 2.1) & 1.55 \\
\bottomrule
\end{tabular}}
\end{table}

The ignition data point the same way. Table~\ref{tab:variants} lists how the fit to each data group changes for rate variants of FFCM-2. Doubling the termination channel, i.e.\ moving it to the measured value, gives the largest improvement and improves the fit in every group, including G4, which was not used to develop FFCM-2. The theoretical rates also improve every group, but by less. Slowing the propagation channel as well worsens the agreement with the \ce{CO2}-diluted data.

\begin{table}[ht]\centering\small
\caption{Screening of \ce{CH3 + HO2} rate variants: change of the file-tempered log-likelihood relative to nominal FFCM-2 (positive: better fit). G1 and G2 contain data used to develop FFCM-2. The RCM set is excluded because of its heat-loss bias.}\label{tab:variants}
\begin{tabular}{lrrrrr}
\toprule
variant & G1 & G2 & G3 & G4 & total \\ \midrule
termination $\times 2.0$ (Hong) & $+3.54$ & $+0.36$ & $+1.83$ & $+0.42$ & $+6.14$ \\
propagation $\times 0.72$ (Hong) & $+3.08$ & $-0.24$ & $+1.33$ & $-0.17$ & $+4.00$ \\
Jasper theory ($\times 1.34$, $\times 0.954$) & $+2.14$ & $+0.26$ & $+1.38$ & $+0.18$ & $+3.96$ \\
termination $\times 3$ & $+3.59$ & $-0.76$ & $-0.09$ & $+0.61$ & $+3.35$ \\
both Hong factors & $+3.97$ & $-1.27$ & $+0.26$ & $+0.25$ & $+3.22$ \\
propagation $\times 1/3$ & $+2.07$ & $-2.82$ & $-0.65$ & $-0.96$ & $-2.35$ \\
\bottomrule
\end{tabular}
\end{table}

\paragraph{Calibration.} The hierarchical calibration quantifies this trend (Fig.~\ref{fig:design}a). Of the four error models compared, the dataset-drift model predicts left-out datasets best (Appendix~\ref{app:calib}): the systematic errors of the datasets vary with temperature, and the high-pressure \ce{CO2}- and argon-diluted data scatter more than their stated uncertainties. With this model, the ignition data alone give a termination factor of 1.51 (90\,\% interval 1.11--1.97) and a propagation factor of 0.76 (0.57--0.97) relative to FFCM-2. Both shifts are in the direction of the measurement, though smaller. Adding the measurement as data does not change these values appreciably, whichever confidence level is assumed for it, so the ignition data carry most of the information.

\paragraph{Robustness.} The individual factors depend on the error model and the prior (Appendix~\ref{app:calib}). The median termination factor stays between 1.3 and 1.6 in all variants, but its interval includes 1 when the publications that supplied targets to FFCM-2 are left out; the evidence for a faster termination channel thus rests largely on data that FFCM-2 was also trained on. The median propagation factor stays below 1, but its interval includes 1 when a laboratory-level offset is added or when FFCM-2's own posterior is used as prior. The most robust result is the direction of the branching: the ratio of the termination factor to the propagation factor lies between 1.5 and 3.4 in all variants, and its 90\,\% interval excludes 1 in every case. The predicted pre-burnt fraction is insensitive to these choices: at the reference condition it lies between 23.7\,\% and 25.1\,\% in all variants.

\paragraph{Structural comparison.} On the observations that both mechanisms can simulate, FFCM-2 predicts left-out datasets better than UoS~sCO$_2$~2.1, but the advantage rests on a few datasets (Appendix~\ref{app:calib}). To fit the data, UoS~sCO$_2$~2.1 must reduce its propagation channel to about a quarter of its value, well below the direct measurement (Fig.~\ref{fig:design}b), and it fits the high-temperature set G4 less well. This may indicate compensation for differences elsewhere in the mechanism. We base the predictions on the calibrated FFCM-2 and include the difference between the two calibrated mechanisms in the uncertainty budget.

\begin{figure}[ht]\centering
\includegraphics[width=\textwidth]{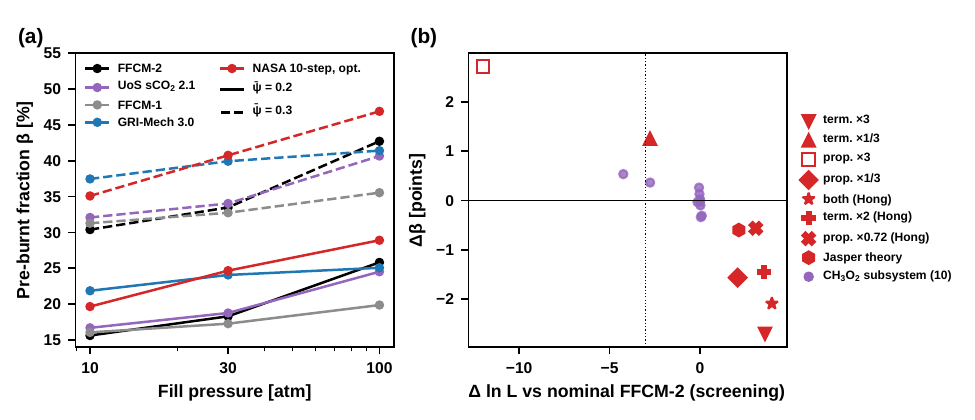}
\caption{(a) Pre-burnt fraction $\beta$ (mixing model~B, $t_\mathrm{fill} = \SI{200}{\micro\second}$) versus fill pressure for the five mechanisms; solid: $\bar\psi = 0.2$, dashed: $\bar\psi = 0.3$. (b) Change of $\beta$ at 100~atm, $\bar\psi = 0.2$ and \SI{200}{\micro\second} for FFCM-2 rate variants versus their change in screening log-likelihood on the argon and nitrogen data (group G1); red: \ce{CH3 + HO2} variants, purple: \cho{} variants. The open square (propagation $\times 3$, $\Delta\ln L = -43.6$) is drawn at the axis limit; dotted line: $\Delta\ln L = -3$.}\label{fig:ensemble}
\end{figure}

\paragraph{Effect on the pre-burnt fraction.} The calibration lowers the predicted pre-burnt fraction by up to about 2 percentage points relative to nominal FFCM-2; at the reference condition, from 25.8\,\% to 24.5\,\% (Fig.~\ref{fig:ensemble}b).

\subsection{Uncertainty budget}\label{sec:budget}
\begin{figure}[ht]\centering
\includegraphics[width=\textwidth]{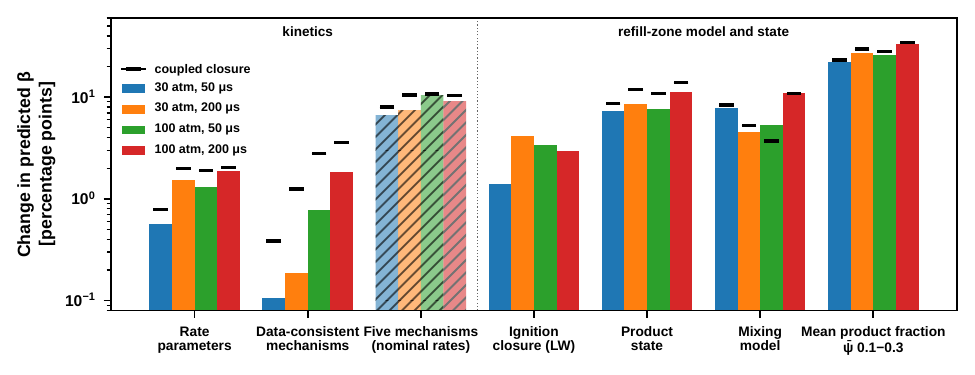}
\caption{Change in the predicted pre-burnt fraction $\beta$ (percentage points, logarithmic axis; mixing model~B) caused by each source of uncertainty at $\bar\psi = 0.2$, for two fill pressures and fill times. Kinetics: rate parameters of the calibrated FFCM-2 (90\,\% range) and the difference between FFCM-2 and UoS~sCO$_2$~2.1 (both calibrated); hatched: the five mechanisms at nominal rates, before calibration. Refill-zone model and state: ignition closure (Livengood--Wu versus coupled integration, calibrated FFCM-2), product state, mixing model, and mean product fraction ($\bar\psi$ from 0.1 to 0.3). Bars: Livengood--Wu closure; black markers: coupled closure.}\label{fig:budget}
\end{figure}

Table~\ref{tab:budget} lists the sources of uncertainty at the reference condition, and Fig.~\ref{fig:budget} shows them at four conditions. With the calibrated mechanisms, two kinetic sources remain: the uncertainty of the rate parameters of FFCM-2, which changes the predicted pre-burnt fraction by about 2 percentage points, and the choice between the two data-consistent mechanisms, which also changes it by about 2 percentage points. Before calibration, the choice among all five mechanisms changed it by 9. The refill-zone sources are larger: the state of the recirculated products and the mixing model each change $\beta$ by about 11 percentage points, five to six times more than either kinetic source, and the mean product fraction changes it by 33 points.

This ranking holds at all 18 conditions with one exception (Appendix~\ref{app:budget}). Over all 18 conditions, the product-state term exceeds the combined kinetic terms by at least a factor of 1.9 and the mixing-model term by at least 1.6, except at 100 atm, $\bar\psi = 0.2$, 50 µs with the coupled closure (0.8). The product-state term exceeds the combined kinetic terms everywhere, by a factor of about 2 or more (at least 2.3 at $\bar\psi = 0.2$). The mixing-model term exceeds them everywhere except at 100~atm and \SI{50}{\micro\second} with the coupled ignition closure, where the two are comparable. A laboratory-level error model widens the term due to the two calibrated factors by a factor of 1.3--1.6, which does not change the ranking. The kinetic terms do not, however, include alternative carbon chemistry under radical seeding, which the data used here do not test directly (Section~\ref{sec:discussion}).

Two modeling choices shift the prediction systematically rather than widening it. The Livengood--Wu closure underestimates $\beta$ relative to the coupled integration of entrainment and chemistry, by about 3 percentage points at the reference condition and by up to 4 at other conditions; the closure error is larger for GRI-Mech~3.0 (up to 8) and negligible for the 10-step mechanism. Treating the fresh propellant as a real gas lowers $\beta$ by about 1 percentage point at 100~atm and by less than 0.2 at 10--30~atm. Both shifts are smaller than the refill-zone terms.

\begin{table}[ht]\centering\small
\caption{Sources of uncertainty in the predicted pre-burnt fraction at the reference condition (100~atm, $\bar\psi = 0.2$, $t_\mathrm{fill} = \SI{200}{\micro\second}$; calibrated FFCM-2, mixing model~B, Livengood--Wu closure: $\beta = 24.5\,\%$). For uncertain parameters the range covers 90\,\% of the calibrated predictions; otherwise it spans the alternatives listed.}\label{tab:budget}
\begin{tabular}{l>{\raggedright\arraybackslash}p{5.6cm}cc}
\toprule
source & what is varied & predicted $\beta$ (\%) & change (points) \\ \midrule
\multicolumn{4}{l}{\emph{Kinetics, calibrated}} \\
rate parameters & two calibrated factors and 256 other FFCM-2 parameters & 23.7--25.5 & 1.9 \\
data-consistent mechanisms & FFCM-2 and UoS~sCO$_2$~2.1, both calibrated & 22.7--24.5 & 1.8 \\
\multicolumn{4}{l}{\emph{Kinetics, before calibration (for comparison)}} \\
five mechanisms & nominal rates & 19.9--28.9 & 9.0 \\
\multicolumn{4}{l}{\emph{Refill zone}} \\
product state & products at \SI{3000}{K}, \SI{2500}{K}, or \SI{3000}{K} without radicals & 13.3--24.5 & 11.2 \\
mixing model & models A and B with alternative parameters & 21.6--32.5 & 10.9 \\
mean product fraction & $\bar\psi$ from 0.1 to 0.3 & 7.8--40.8 & 33.0 \\
\multicolumn{4}{l}{\emph{Systematic shifts}} \\
ignition closure & coupled integration instead of Livengood--Wu & 25.0 $\to$ 28.0$^a$ & $+3.0$ \\
real-gas fresh propellant & Peng--Robinson enthalpy at \SI{300}{K} & 24.5 $\to$ 23.6 & $-0.9$ \\
\bottomrule
\end{tabular}\\[2pt]
{\footnotesize $^a$Both values on the same grid of entrainment rates; the Livengood--Wu value on this grid is 25.0\,\%,
slightly above the 24.5\,\% of the main calculation.}
\end{table}

\subsection{Value of additional experiments}
\begin{figure}[ht]\centering
\includegraphics[width=\textwidth]{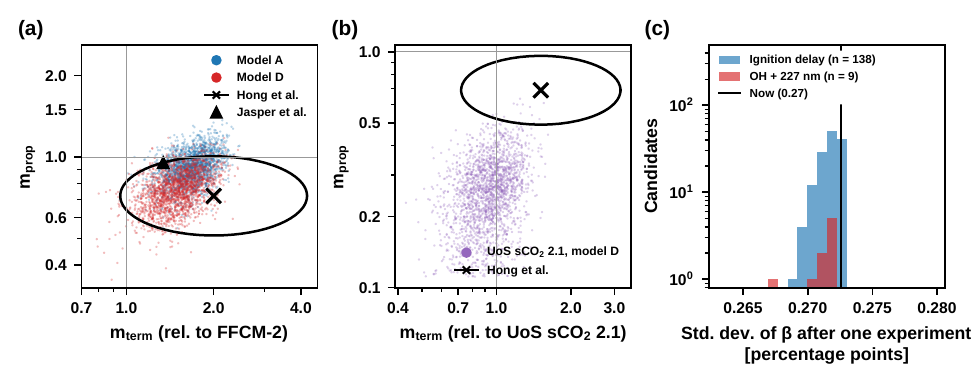}
\caption{(a) Calibrated \ce{CH3 + HO2} rate factors for FFCM-2 from ignition data alone, under the error model with dataset offsets only (model~A) and the dataset-drift model (model~D), with the measurement of \cite{Hong2012} (cross and two-standard-deviation ellipse) and the theory of \cite{Jasper2009} (triangle). (b) The same for UoS~sCO$_2$~2.1 (dataset-drift model, 379 common observations), with the measurement expressed relative to that mechanism. (c) Number of candidate experiments (logarithmic scale) by the remaining standard deviation of $\beta$ (root mean square over 18 conditions) after one experiment; vertical line: current value.}\label{fig:design}
\end{figure}

The design analysis addresses only the remaining uncertainty of the two calibrated rate factors. The other FFCM-2 parameters, the choice between the calibrated mechanisms and the untested carbon chemistry under radical seeding are outside its scope, and the candidate set contains no radical-seeded experiments. With the dataset-drift model, the remaining uncertainty corresponds to a standard deviation of $\beta$ of 0.27 percentage points (root mean square over the 18 conditions) (Table~\ref{tab:design}, Fig.~\ref{fig:design}c).

Ignition-delay experiments barely reduce this uncertainty: the best series of four in one apparatus reduces it by about 1--2\,\%, because the sensitivity of the ignition delay to the branching is small compared with the offset and drift expected for a new dataset. Rate experiments of the type of \cite{Hong2012} help somewhat more: the best series of four reduces the uncertainty by about 3\,\%. Their signals are dominated by \ce{H2O2} decomposition and by the uncertainty of the shock temperature, and at 30~atm many late-time \ce{OH} and \ce{HO2} signals fall below the detection limit. This noise model is plausible: for a series of experiments at Hong et al.'s conditions, it reproduces the uncertainty they reported if their factors are read as two standard deviations. With a laboratory-level error model, the starting uncertainty is larger (0.38 percentage points), and the best rate series reduces it by about 20\,\%; ignition-delay experiments still do not help. The value of rate experiments also depends on the noise assumptions (Appendix~\ref{app:design}).

\begin{table}[ht]\centering\small
\caption{Value of additional experiments: standard deviation of $\beta$ due to the two calibrated rate factors (percentage
points, root mean square over 18 conditions) before and after the best experiments, for the dataset-drift error model and its
laboratory-level variant. A series consists of four experiments in one apparatus.}\label{tab:design}
\begin{tabular}{>{\raggedright\arraybackslash}p{8.2cm}cc}
\toprule
design & dataset-drift & laboratory-level \\
 & model & model \\ \midrule
current (ignition data only) & 0.273 & 0.379 \\
best series of ignition-delay experiments (\ce{CO2}-diluted) & 0.269 & 0.374 \\
best single rate experiment (\ce{H2O2}/\ce{CH4}/Ar, 3.5 atm, 1050 K) & 0.267 & 0.338 \\
best series of rate experiments (3.5--30 atm, 1050--1250 K) & 0.264 & 0.305 \\
\bottomrule
\end{tabular}
\end{table}

Within this framework, the best kinetic experiments considered here would reduce the remaining parametric uncertainty by 3--20\,\%, depending on the error model. The refill-zone sources are larger, but the measurements that would constrain them were not evaluated with a noise model here. The comparison therefore indicates priorities under the stated assumptions rather than the value of a particular measurement.

\section{Discussion}\label{sec:discussion}
\paragraph{Implications for RDE simulations.} Mechanisms without \cho{} chemistry ignite too slowly when compared with the high-pressure low-temperature laboratory data (Table~\ref{tab:audit}). For GRI-Mech~3.0 the sign of the error depends on temperature: it ignites too slowly below about \SI{1300}{K} but about 1.4 times too fast above, at $\geq 40$~atm. Its error in pre-ignition estimates for fresh propellant at mixing temperatures of 1000--2000~K, such as those of \cite{Stechmann2019}, therefore depends on the temperature range considered.

In refill mixtures the comparison changes. GRI-Mech~3.0 predicts from about 1 percentage point less to 8 points more pre-burning than FFCM-2, because it ignites quickly in radical-seeded, \ce{H2O}-rich mixtures; with radical-free products the difference falls to at most about 3 points. Two hybrid mechanisms locate the cause (Appendix~\ref{app:audit}). Replacing the \ce{H2}/\ce{O2} subsystem of FFCM-2 with that of GRI-Mech~3.0 changes neither the agreement with the ignition data nor $\beta$ appreciably, so the difference is presumed to lie in the carbon chemistry of GRI-Mech~3.0. Adding the \cho{} subsystem of FFCM-2 to GRI-Mech~3.0 yields a mechanism that ignites 1.3--2.2 times too fast in the shock-tube data and is therefore excluded; a retuned combination, however, cannot be ruled out. At the temperatures that determine $\beta$ at 30--100~atm (about 930--1290~K near the critical product fraction), GRI-Mech~3.0 ignites too slowly on its own and too fast with the added \cho{} subsystem; its over-reactivity above \SI{1300}{K} matters mainly at 10~atm and short fill times.

The NASA 10-step mechanism, which lacks \ce{CH3}, \ce{HO2} and \cho{}, reproduces neither the laboratory data nor the detailed mechanisms in refill mixtures, where it predicts the most pre-burning at high pressure. Our analysis suggests that reduced mechanisms for RDE refill-zone chemistry therefore need to retain the \ce{CH3}/\ce{HO2}/\cho{} subsystem. They should be assessed by the pre-burnt fraction they predict, relative to the uncertainty established here, and not only by ignition delays at standard conditions.

\paragraph{Implications for kinetics.} The most robust kinetic result is the direction of the \ce{CH3 + HO2} branching. The ignition data favor a larger share of the termination channel than FFCM-2 assumes, consistent with the direct measurement of \cite{Hong2012}; with FFCM-2's own posterior as prior, the result is also consistent with theory \cite{Jasper2009}. The individual channels are less certain. Relative to theory, the calibrated termination channel is only about 13\,\% faster, well within the theoretical uncertainty, and only the propagation channel lies clearly below theory. The evidence for a faster termination channel rests largely on data that FFCM-2 was also trained on, and the propagation factor cannot be separated from a common offset of all shock-tube data. We are not proposing new rate coefficients. Because the calibration varies only these two channels, the factors may compensate for deficiencies elsewhere in the mechanism.  The errors of FFCM-2’s low-temperature pathway also differ in sign between datasets, too slow in \cite{Huang2004} at 1000–1200 K and too fast in \cite{Burke2015,Mathieu2015} near 1200–1300 K, which constant rate factors can compensate only on average. A direct measurement of the branching at elevated pressure would test the calibrated factors, but according to our design analysis it would change the predicted pre-burnt fraction only slightly.

\paragraph{Implications for experiments.} Within our framework, the chemistry no longer appears to be the leading uncertainty in predicted pre-burning. With the calibrated mechanisms, the kinetic sources change the predicted pre-burnt fraction by about 2 percentage points each at the reference condition, and the best kinetic experiments considered would reduce the parametric part only slightly. The state of the recirculated products and the mixing model each change it by about 11 points. Because the refill-zone descriptions are themselves idealized, and the measurements that would constrain them were not evaluated with a noise model, this comparison indicates priorities rather than the value of a specific measurement. Measurements of the temperature and composition of the recirculated products would address the largest refill-zone source. Rotating detonation engines in which the degree of reactant premixing can be varied \cite{Ayers2023} offer one way to measure how the mixing state affects pre-burning. For the kinetics, experiments in radical-seeded, product-diluted mixtures at elevated pressure would test the carbon chemistry that the present data leave open.

\paragraph{Limitations of the chemistry and data analysis.} Beyond the approximations stated in the Introduction, the following limitations apply. The random dataset effects absorb constant and temperature-linear systematic errors, but not more complex ones. The carbon chemistry of GRI-Mech~3.0 under radical seeding, which explains much of its difference from FFCM-2 in refill mixtures, is tested by the data used here only without radical seeding. Mixing rules for multi-component colliders, which can change effective rates by up to a factor of 2.3 in \ce{Ar}/\ce{CO2} mixtures \cite{ShaoH2O2,BurkeSong2017}, are not represented in any of the mechanisms, although they are relevant to the \ce{H2O}- and \ce{CO2}-rich refill gas. The RCM data lack heat-loss histories and are used only qualitatively. The ignition data are simulated with ideal-gas thermodynamics, which makes the most compressed \ce{CO2}-diluted mixtures a few percent too dense in the simulation. Further details on data selection are given in Appendix~\ref{app:audit}.

\paragraph{Limitations of the refill-zone and cycle models.} The cycle estimate uses ideal-gas thermodynamics at states far from ideal. At the Chapman--Jouguet state (4190--4670~K, 320--3450~atm), the compressibility factor is about 1.03, 1.08 and 1.27 for fill pressures of 10, 30 and 100~atm, and the thermodynamic fits of the product species are extrapolated beyond their upper limit of \SI{3500}{K}. The ideal gains, their decrease with fill pressure and the effect of a reacted fraction on the Chapman--Jouguet pressure are therefore indicative, particularly at 100~atm. The refill mixtures themselves are close to ideal, but the non-ideal enthalpy of the fresh propellant lowers $\beta$ by about 1 percentage point at 100~atm (Section~\ref{sec:budget}). The Livengood--Wu closure of the progressive-entrainment model underestimates $\beta$ by up to 4 percentage points and should be replaced by coupled integration in future work. A spatially resolved treatment of the refill zone, with directly computed rather than prescribed mixing and product states, would be the next research step.

\section{Conclusions}
Within an idealized, zero-dimensional description of the refill zone, this study shows the following.
\begin{enumerate}
\item The choice of kinetic mechanism changes the predicted refill-zone ignition delay by up to a factor of about 20 and the predicted pre-burnt fraction by up to 13 percentage points. In the ideal cycle, however, this corresponds to at most about one percentage point of specific-impulse gain; the pre-burnt fraction matters mainly through effects outside the ideal cycle, such as changes of the detonation mode.
\item Mechanisms without methylperoxy chemistry, including the NASA 10-step mechanism, do not reproduce the available high-pressure ignition data below about \SI{1150}{K}, even when a facility pressure rise is allowed for. The evidence comes from a few independent sources, and near 1200--1300~K the low-temperature pathway of FFCM-2 appears too fast. 
\item Calibration against the ignition data favors a larger share of the chain-terminating channel of \ce{CH3 + HO2}, consistent with a direct rate measurement. The individual rate factors depend on the error model and prior and should be regarded as effective values.
\item With the calibrated mechanisms, the remaining kinetic uncertainty in the predicted pre-burnt fraction is smaller, in most cases several times smaller, than the uncertainty arising from the state of the recirculated products and the mixing model, although it does not cover alternative carbon chemistry under radical seeding. The best kinetic experiments considered would reduce its parametric part only slightly. Characterizing the refill zone should therefore take priority.
\end{enumerate}

\section*{Data availability}
The ChemKED database and the FFCM-2 posterior and validation table are publicly available from their developers. The UoS~sCO$_2$~2.1 mechanism is distributed with \cite{HarmanThomasThesis}.

\appendix
\section{Calibration details}\label{app:calib}
\paragraph{Surrogates and sampling.} The quadratic surrogates of $\ln\tau$ were validated on four additional grid points per observation (largest error 0.048 in $\ln\tau$, median 0.010). The surrogates of $\beta$ were fitted on local $3\times3$ grids around each posterior (half-width factor 1.4 for FFCM-2 and 1.8 for UoS~sCO$_2$~2.1; validation errors at most 0.07 percentage points). The random dataset effects are integrated out analytically with the Woodbury identity. Each fit uses four adaptive-Metropolis chains; in every fit reported, the split-$\hat R$ statistic is at most 1.009 and the effective sample size at least 430.

\paragraph{Error models.} Four error models of increasing generality were compared: dataset offsets only (model~A); offsets and drifts in inverse temperature (B); offsets and extra scatter per data group (C); and offsets, drifts and group scatter (D, the dataset-drift model of the main text). They are compared by the expected log predictive density for left-out datasets (elpd, in natural-logarithm units), estimated by Pareto-smoothed importance sampling (PSIS-LOO) \cite{Vehtari2017}. Model~D improves on model~A by 66 (standard error 24), on model~B by 41 (25) and on model~C by 30 (13); without the datasets for which the Pareto diagnostic $\hat k$ exceeds 0.7, the improvements are 44 (14), 17 (5) and 21 (11). Two robustness variants below (a common offset, a wider drift prior) score only marginally higher than model~D, by 0.8 (2.0) and 1.8 (2.4). The fitted drift scale corresponds to about 1.6 in $\ln\tau$ per unit of $1000/T$.

\paragraph{Robustness variants.} Five variants test the assumptions of model~D:
(i) a common offset and drift shared by all shock tubes, because a systematic error such as the neglected facility pressure rise need not average to zero;
(ii) an additional offset shared by all datasets of one laboratory (Stanford: Petersen et al.\ and Shao et al.; UCF: Pryor et al.\ and Barak et al.; Shao~2019 split by facility);
(iii) a four times wider prior on the drift scale;
(iv) the publications that supplied methane ignition targets to FFCM-2 (Petersen et al.\ 1996 and 1999, Shao et al.\ 2018 and 2019) left out, so that no observation is used twice; and
(v) FFCM-2's own posterior for the two channels as prior.
Table~\ref{tab:posterior} gives the results. With a laboratory-level offset, the laboratory scale (0.47) exceeds the dataset scale (0.18), so the effective number of independent units is closer to the 10 laboratories than to the 90 datasets. Leave-one-out estimates at the laboratory level are not reported, because importance sampling fails when a whole laboratory is left out ($\hat k > 1$). 

With a common offset (estimated at $-0.14$), the propagation factor is only weakly constrained and reaches the edge of the surrogate domain.  This is because the offset and a slower propagation channel compensate each other; this variant improves the elpd by only 0.8 (standard error 2.0). FFCM-2's own posterior as prior partly double-counts the data of Petersen et al.\ and Shao et al., so the defensible estimate lies between this variant and model~D. The ratio of the termination to the propagation factor is 1.98 (1.48--2.62) for model~D, 2.20 (1.45--3.62) with a laboratory offset, 2.56 (1.45--4.01) without the FFCM-2 target sources, 1.49 (1.18--1.88) with FFCM-2's prior and 3.42 (1.94--5.31) with a common offset.

\paragraph{Direct measurement as data.} Hong et al.\ \cite{Hong2012} do not state a confidence level. Their uncertainty factors (1.4 and 2.1) are root-sum-square combinations of component limits, partly taken from the evaluation of Baulch et al., whose limits are commonly interpreted as two or three standard deviations \cite{Nagy2015}. Adding the measurement to model~D gives termination factors of 1.53, 1.57 and 1.63, and propagation factors of 0.76, 0.75 and 0.75, for readings as one, two and three standard deviations.

\paragraph{Structural comparison.} On the 379 observations that both mechanisms can simulate (the UoS mechanism contains no He), FFCM-2 predicts left-out datasets better by 35 (standard error 22). Without the three datasets with $\hat k > 0.7$, the difference is 10 (standard error 9).

\begin{table}[t]\centering\small
\caption{Hierarchical calibration of the \ce{CH3 + HO2} channels from ignition data alone: medians (90\,\% intervals) of the rate factors relative to each mechanism, and the expected log predictive density for left-out datasets (elpd, with standard error; higher is better; comparable only on the same observations).}\label{tab:posterior}
\resizebox{\textwidth}{!}{%
\begin{tabular}{llcccr}
\toprule
mechanism & error model & observations & $m_\mathrm{term}$ & $m_\mathrm{prop}$ & elpd \\ \midrule
FFCM-2 & A: dataset offsets & 402 & 1.67 (1.32--2.06) & 0.91 (0.73--1.12) & 20.5 (40.7) \\
FFCM-2 & B: + drift in $1000/T$ & 402 & 1.39 (1.04--1.81) & 0.77 (0.58--0.98) & 44.7 (40.1) \\
FFCM-2 & C: + group scatter & 402 & 1.36 (1.07--1.68) & 0.72 (0.55--0.91) & 55.9 (36.2) \\
FFCM-2 & D: drift + group scatter & 402 & 1.51 (1.11--1.97) & 0.76 (0.57--0.97) & 86.2 (31.3) \\
\midrule
FFCM-2 & D + laboratory offset & 402 & 1.47 (1.10--1.90) & 0.66 (0.38--1.06) & -- \\
FFCM-2 & D + common offset and drift & 402 & 1.57 (1.05--2.21) & 0.44 (0.34--0.77) & 87.0 (31.8) \\
FFCM-2 & D, wider prior on drift scale & 402 & 1.56 (1.13--2.10) & 0.77 (0.59--0.98) & 88.0 (30.4) \\
FFCM-2 & D, FFCM-2's posterior as prior & 402 & 1.29 (1.00--1.62) & 0.86 (0.72--1.02) & 85.2 (30.4) \\
FFCM-2 & D, without FFCM-2 target sources & 191 & 1.48 (0.80--2.42) & 0.58 (0.36--0.94) & -- \\
\midrule
FFCM-2 & D, common observations & 379 & 1.71 (1.26--2.23) & 0.74 (0.53--0.96) & 85.4 (32.0) \\
UoS sCO$_2$ 2.1 & D, common observations & 379 & 0.94 (0.65--1.26) & 0.26 (0.13--0.41) & 49.9 (26.0) \\
\bottomrule
\end{tabular}}
\end{table}

\section{Comparison with experimental data: details}\label{app:audit}
\paragraph{Data selection.} Seven ChemKED observations, all from one 1975 study, are excluded because they do not ignite in the simulation at any surrogate grid point; their exclusion does not depend on the rate factors. The temperature limit $T_5 \leq \SI{1500}{K}$ applies to G1 only. G4 (1440--1855~K) is the one high-temperature set not used to develop FFCM-2 and is retained in full as a check outside the refill-relevant range. At \SI{1100}{K} and 286~atm, the compressibility factor of \ce{CO2} is about 1.06, so the simulated concentrations of the most compressed \ce{CO2}-diluted mixtures are a few percent too high and their simulated delays slightly short.

\paragraph{Discriminating points.} Of the 775 ChemKED points simulated by all five detailed mechanisms, 136 discriminate between FFCM-2 and its \cho{}-free variant (difference larger than twice the stated uncertainty). Of these, 110 favor FFCM-2 and 26 the \cho{}-free variant. Of the 110, 96 come from \cite{Petersen1999JPP}, which also contributes 15 of the 26, and 14 from \cite{Huang2004} (1000--1200~K, 30--39~atm), where FFCM-2 itself ignites 1.2--2.9 times too slowly and its \cho{}-free variant 2.2--18 times too slowly. All 11 discriminating points of \cite{Burke2015,Mathieu2015} (25--48~atm, 1190--1330~K, \ce{N2} dilution) favor the \cho{}-free variant, with FFCM-2 1.2--1.8 times too fast. On the 123 points at $\geq 10$~atm and $\leq \SI{1500}{K}$ from publications that did not supply FFCM-2 targets, the rms of $\ln(\tau_\mathrm{exp}/\tau_\mathrm{model})$ is 0.54 for FFCM-2, 0.68 for UoS~sCO$_2$~2.1, 0.83 for FFCM-2 without \cho{}, 0.84 for FFCM-1 and 0.86 for GRI-Mech~3.0. The five of these points in the \cho{}-sensitive region favor the \cho{}-free mechanisms (rms 0.16 for FFCM-2 without \cho{}, 0.20 for FFCM-1 and 0.53 for FFCM-2).

\paragraph{Facility pressure rise.} The 311 G1 calibration points were re-simulated with FFCM-2 and its \cho{}-free variant under a pressure rise of 2 or 4\,\% per ms, imposed as isentropic compression. This shortens FFCM-2's delays by up to 14 or 23\,\% (median 1--2\,\%). In the \cho{}-sensitive region, the mean $\ln(\tau_\mathrm{exp}/\tau_\mathrm{model})$ of the \cho{}-free variant changes from $-0.80$ to $-0.74$ at 4\,\% per ms, that of FFCM-2 from $+0.10$ to $+0.12$, and the number of discriminating points changes by at most three.

\paragraph{NASA 10-step mechanism.} On the ChemKED points where it ignites, the rms of $\ln(\tau_\mathrm{exp}/\tau_\mathrm{model})$ is 1.9; it does not ignite within 20 measured delays at 241 of the 775 points.

\paragraph{Hybrid mechanisms.} GRI-Mech~3.0 predicts between 1.3 percentage points less and 7.8 more pre-burning than nominal FFCM-2 over the 18 conditions (0.1 to 8.5 more than the calibrated FFCM-2); with radical-free products, at most 2.7 more. FFCM-2 with its 29 \ce{H2}/\ce{O2} reactions replaced by the 30 of GRI-Mech~3.0 reproduces the ignition data as well as FFCM-2 (rms 0.37, 0.30, 0.21 and 0.12 in G1--G4, versus 0.37, 0.37, 0.25 and 0.14) and changes $\beta$ by $-1.9$ to $+0.9$ percentage points ($-1.3$ to $+1.2$ with the calibrated factors). GRI-Mech~3.0 with the 33 \cho{} reactions of FFCM-2 ignites 1.3--2.2 times too fast in G1--G4 (rms 0.67 in G1, versus 0.37 for FFCM-2) and would predict 0.5--8.9 percentage points more pre-burning than nominal FFCM-2. Near the critical product fraction, the mixing temperature is about 930--1290~K at 30--100~atm; at $\geq 40$~atm and below \SI{1300}{K}, GRI-Mech~3.0 alone ignites about 1.35 times too slowly and with the added \cho{} subsystem about 1.8 times too fast. Above \SI{1300}{K}, it ignites about 1.4 times too fast in G1 at $\geq 40$~atm and 1.3 times too fast in G4.

\section{Uncertainty budget: details}\label{app:budget}
\paragraph{Ranges over all conditions.} Over the 18 conditions, the rate-parameter term (90\,\% range of 60 joint samples of the two calibrated factors and the other 256 FFCM-2 parameters) is 0.1--2.9 percentage points; with only the two calibrated factors, 0.1--1.8. The other parameters are sampled from FFCM-2's covariance conditional on the two calibrated channels; because that covariance is overconfident in this regime (Section~\ref{sec:audit}), the term is probably underestimated. With 60 samples the range itself is uncertain; at 100~atm, $\bar\psi = 0.3$ and \SI{200}{\micro\second}, for example, it is 2.9 points with a bootstrap 90\,\% interval of 2.3--3.2. The difference between the calibrated FFCM-2 and UoS~sCO$_2$~2.1 is 0.0--2.8 points. At $\bar\psi = 0.2$ and 30--100~atm, the product state contributes 7.3--11.2 points, the mixing model 4.6--10.9 and the mean product fraction 22--33, and the smaller of the product-state and mixing-model terms exceeds the sum of the two kinetic terms by a factor of 2.5--11. A laboratory-level error model widens the term of the two calibrated factors from 0.5--1.2 to 0.8--1.6 points.

\paragraph{Coupled ignition closure.} Relative to coupled integration, the Livengood--Wu closure underestimates $\beta$ by 0.3--4.1 percentage points for the calibrated FFCM-2, by up to 8.1 for GRI-Mech~3.0, and hardly at all for the 10-step mechanism; the spread among the five mechanisms becomes 3.0--13.3 points instead of 1.0--12.8. Under the coupled closure at $\bar\psi = 0.2$ and 30--100~atm, the two calibrated factors contribute 0.7--1.6 points (1.1--1.4 times the Livengood--Wu value; central differences with factor-1.5 steps), the full rate-parameter term 0.8--2.1, the two mechanisms 0.4--3.6, the product state 8.7--13.9, the mixing model 3.7--10.9 and the mean product fraction 23--34. The product state then exceeds the combined kinetic terms by a factor of 2.3--7.5 and the mixing model by 0.8--7.2. Halving the operator-splitting step changes the sensitivities of $\beta$ to the rate factors by a median of 1--9\,\% and at most 22\,\% at 10, 30 and 100~atm.

\paragraph{Specific impulse.} Across the five mechanisms, each percentage point of $\beta$ lowers the ideal impulse gain by 0.020--0.028 percentage points in the parcel model, 0.083--0.091 if pre-burnt pockets stay segregated, and 0.03--0.17 if they mix into the charge (largest at $\bar\psi = 0.1$). Matched condition by condition, the spread of $\beta$ among the five mechanisms corresponds to 0.16--0.59 percentage points of ideal gain in the homogeneous limit and 0.09--1.07 in the segregated limit. At $\bar\psi = 0.2$, keeping the gain within $\pm 0.5$ percentage points tolerates errors in $\beta$ of about $\pm 6$ (segregated) to $\pm 7$--10 (homogeneous) percentage points. In the homogeneous limit, a reacted fraction of 5\,\% lowers the Chapman--Jouguet pressure ratio at 30~atm from 33 to 18.

\paragraph{Real-gas effects.} With a Peng--Robinson equation of state, the fresh propellant at \SI{300}{K} has a compressibility factor of 0.987, 0.964 and 0.905 and an enthalpy departure of $-4.6$, $-14$ and $-43$~kJ/kg at 10, 30 and 100~atm. This lowers the mixing temperature by about 2, 6 and 19~K and $\beta$ of the calibrated FFCM-2 by up to 0.06, 0.2 and 1.6 percentage points (0.3--0.9 at 100~atm and $\bar\psi = 0.2$). At the Chapman--Jouguet states, a co-volume estimate gives compressibility factors of 1.03, 1.08 and 1.27, and the thermodynamic fits of the product species, valid to \SI{3500}{K}, are extrapolated by 700--1200~K.

\section{Experimental design: details}\label{app:design}
\paragraph{Candidates.} The 138 ignition-delay candidates use \ce{CH4}/\ce{O2} with Ar at $\phi = 0.5$, 1.1 and 2, with 70\,\% \ce{CO2}, or with radical-free product species, at \SIrange{950}{1300}{K} and \SIrange{10}{100}{atm}, and delays between \SI{10}{\micro\second} and \SI{100}{ms} (delays above \SI{10}{ms} only in an RCM at $\leq \SI{1100}{K}$). The nine rate experiments record \ce{OH} and the 227-nm absorbance ($\propto 13.1\,X_{\ce{HO2}} + 1.4\,X_{\ce{H2O2}}$, with the cross sections of \cite{Hong2012}) at 20--200~\si{\micro\second}. Signals below an assumed detection limit (2~ppm \ce{OH} and 20~ppm \ce{HO2}-equivalent at 3.5~atm, scaled with $1/p$) are discarded. Each candidate observable has its own quadratic surrogate, whose validation error is added to the noise.

\paragraph{Noise model.} For ignition delays, the noise comprises the stated measurement uncertainty (0.09 in $\ln\tau$) and an offset, drift and scatter for the new apparatus drawn from the calibrated error model. For rate experiments, it comprises per-point noise of 0.05 (\ce{OH}) and 0.10 (227~nm); separate calibration biases of 0.10, shared by a series; uncertainties of 0.7\,\% in the reflected-shock temperature and 5\,\% in the initial \ce{H2O2} loading for each experiment; and the uncertainty of all other FFCM-2 rates, shared by every experiment of a series. \ce{H2O2} decomposition (its rate and argon collision efficiency) carries 81--100\,\% of the variance due to the other rates. Each expected utility is estimated five times with independent outcome draws; all Monte Carlo standard errors are below 0.001 percentage points.

\paragraph{Plausibility of the noise model.} For a series of 6--12 experiments at the conditions of \cite{Hong2012}, the noise model gives uncertainty factors (two standard deviations) of 2.2--1.8 for the termination channel and 1.5--1.4 for the propagation channel. These are close to the factors Hong et al.\ reported (2.1 and 1.4) if those are read as two standard deviations; read as one standard deviation, their uncertainty would be larger and our noise model optimistic.

\paragraph{Sensitivity to the noise assumptions.} For the dataset-drift model, the best series of rate experiments reduces the standard deviation of $\beta$ from 0.273 to 0.255 percentage points without the detection limit, to 0.242 if the other rates, the temperature and the \ce{H2O2} loading are known exactly, and to 0.199 with neither restriction and without surrogate error.

\bibliographystyle{unsrtnat}
\bibliography{refs}

@article{Hashemi2016,
  author  = {Hashemi, H. and Christensen, J. M. and Gersen, S. and Levinsky, H. and Klippenstein, S. J. and Glarborg, P.},
  title   = {High-pressure oxidation of methane},
  journal = {Combust. Flame}, volume = {172}, pages = {349--364}, year = {2016},
  doi     = {10.1016/j.combustflame.2016.07.016}}

@article{HarmanThomas2023,
  author  = {Harman-Thomas, J. M. and Ingham, D. B. and Hughes, K. J. and Pourkashanian, M.},
  title   = {Role of methyldioxy radical chemistry in high-pressure methane combustion in {CO$_2$}},
  journal = {Int. J. Chem. Kinet.}, volume = {55}, pages = {629--641}, year = {2023}, doi = {10.1002/kin.21672}}

@article{HarmanThomas2022,
  author  = {Harman-Thomas, J. M. and Hughes, K. J. and Pourkashanian, M.},
  title   = {The development of a chemical kinetic mechanism for combustion in supercritical carbon dioxide},
  journal = {Energy}, volume = {255}, pages = {124490}, year = {2022}, doi = {10.1016/j.energy.2022.124490}}

@phdthesis{HarmanThomasThesis,
  author  = {Harman-Thomas, J. M.},
  title   = {Shock Tube Design and Fabrication for the Development of a Kinetic Model for Supercritical {CO$_2$} Production},
  school  = {University of Sheffield}, type = {{EngD} thesis}, year = {2023},
  note    = {https://etheses.whiterose.ac.uk/id/eprint/33138/}}

@article{Karimi2019,
  author  = {Karimi, M. and Ochs, B. and Liu, Z. and Ranjan, D. and Sun, W.},
  title   = {Measurement of methane autoignition delays in carbon dioxide and argon diluents at high pressure conditions},
  journal = {Combust. Flame}, volume = {204}, pages = {304--319}, year = {2019}, doi = {10.1016/j.combustflame.2019.03.020}}

@article{Shao2019,
  author  = {Shao, J. and Choudhary, R. and Davidson, D. F. and Hanson, R. K. and Barak, S. and Vasu, S.},
  title   = {Ignition delay times of methane and hydrogen highly diluted in carbon dioxide at high pressures up to 300 atm},
  journal = {Proc. Combust. Inst.}, volume = {37}, pages = {4555--4562}, year = {2019}, doi = {10.1016/j.proci.2018.08.002}}

@article{Shao2020,
  author  = {Shao, J. and Ferris, A. M. and Choudhary, R. and Cassady, S. J. and Davidson, D. F. and Hanson, R. K.},
  title   = {Shock-induced ignition and pyrolysis of high-pressure methane and natural gas mixtures},
  journal = {Combust. Flame}, volume = {221}, pages = {364--370}, year = {2020}}

@article{Barak2020,
  author  = {Barak, S. and Pryor, O. and Ninnemann, E. and Neupane, S. and Vasu, S. and Lu, X. and Forrest, B.},
  title   = {Ignition delay times of oxy-syngas and oxy-methane in supercritical {CO$_2$} mixtures for direct-fired cycles},
  journal = {J. Eng. Gas Turbines Power}, volume = {142}, pages = {021014}, year = {2020}, doi = {10.1115/1.4045743}}

@article{Pryor2017JERT,
  author  = {Pryor, O. and Barak, S. and Lopez, J. and Ninnemann, E. and Koroglu, B. and Nash, L. and Vasu, S. S.},
  title   = {High pressure shock tube ignition delay time measurements during oxy-methane combustion with high levels of {CO$_2$} dilution},
  journal = {J. Energy Resour. Technol.}, volume = {139}, pages = {042208}, year = {2017}}

@article{Petersen1999CF,
  author  = {Petersen, E. L. and Davidson, D. F. and Hanson, R. K.},
  title   = {Kinetics modeling of shock-induced ignition in low-dilution {CH$_4$/O$_2$} mixtures at high pressures and intermediate temperatures},
  journal = {Combust. Flame}, volume = {117}, pages = {272--290}, year = {1999}}

@article{Petersen1999JPP,
  author  = {Petersen, E. L. and Davidson, D. F. and Hanson, R. K.},
  title   = {Ignition delay times of ram accelerator {CH$_4$/O$_2$}/diluent mixtures},
  journal = {J. Propuls. Power}, volume = {15}, pages = {82--91}, year = {1999}}

@article{Hong2012,
  author  = {Hong, Z. and Davidson, D. F. and Lam, K.-Y. and Hanson, R. K.},
  title   = {A shock tube study of the rate constants of {HO$_2$} and {CH$_3$} reactions},
  journal = {Combust. Flame}, volume = {159}, pages = {3007--3013}, year = {2012}}

@article{Jasper2009,
  author  = {Jasper, A. W. and Klippenstein, S. J. and Harding, L. B.},
  title   = {Theoretical rate coefficients for the reaction of methyl radical with hydroperoxyl radical and for methylhydroperoxide decomposition},
  journal = {Proc. Combust. Inst.}, volume = {32}, pages = {279--286}, year = {2009}}

@article{Xu2022,
  author  = {Xu, S. and Liang, J. and Cao, S. and He, R. and Yin, G. and Wang, Q.-D.},
  title   = {A hierarchical theoretical study of the hydrogen abstraction reactions of {H$_2$/C$_1$--C$_4$} molecules by the methyl peroxy radical and implications for kinetic modeling},
  journal = {ACS Omega}, volume = {7}, pages = {8675--8685}, year = {2022}}

@article{Tsang1986,
  author  = {Tsang, W. and Hampson, R. F.},
  title   = {Chemical kinetic data base for combustion chemistry. {Part I}. {Methane} and related compounds},
  journal = {J. Phys. Chem. Ref. Data}, volume = {15}, pages = {1087--1279}, year = {1986}}

@article{BurkeSong2017,
  author  = {Burke, M. P. and Song, R.},
  title   = {Evaluating mixture rules for multi-component pressure dependence: {H + O$_2$ (+M) = HO$_2$ (+M)}},
  journal = {Proc. Combust. Inst.}, volume = {36}, pages = {245--253}, year = {2017}}

@article{ShaoH2O2,
  author  = {Shao, J. and Choudhary, R. and Davidson, D. F. and Hanson, R. K.},
  title   = {Shock tube/laser absorption measurement of the rate constant of the reaction: {H$_2$O$_2$ + CO$_2$ $\to$ 2OH + CO$_2$}},
  journal = {Proc. Combust. Inst.}, volume = {39}, number = {1}, pages = {735--743}, year = {2023}, doi = {10.1016/j.proci.2022.08.021}}

@misc{GRI30,
  author  = {Smith, G. P. and Golden, D. M. and Frenklach, M. and Moriarty, N. W. and Eiteneer, B. and Goldenberg, M. and Bowman, C. T. and Hanson, R. K. and Song, S. and Gardiner, W. C. and Lissianski, V. V. and Qin, Z.},
  title   = {{GRI-Mech} 3.0}, year = {1999}, note = {http://combustion.berkeley.edu/gri-mech/}}

@misc{FFCM1,
  author  = {Smith, G. P. and Tao, Y. and Wang, H.},
  title   = {Foundational Fuel Chemistry Model Version 1.0 ({FFCM-1})}, year = {2016},
  note    = {http://nanoenergy.stanford.edu/ffcm1}}

@misc{FFCM2,
  author  = {Zhang, Y. and Dong, W. and Vandewalle, L. A. and Xu, R. and Smith, G. P. and Wang, H.},
  title   = {Foundational Fuel Chemistry Model Version 2.0 ({FFCM-2})}, year = {2023},
  note    = {\url{https://web.stanford.edu/group/haiwanglab/FFCM2} (mechanism, posterior covariance and targets used here from
             \url{https://github.com/yzhang9312/FFCM22023})}}

@article{Zhang2024,
  author  = {Zhang, Y. and Dong, W. and Xu, R. and Smith, G. P. and Wang, H.},
  title   = {Foundational Fuel Chemistry Model 2 -- iso-Butene chemistry and application in modeling alcohol-to-jet fuel combustion},
  journal = {Combust. Flame}, volume = {259}, pages = {113168}, year = {2024}, doi = {10.1016/j.combustflame.2023.113168}}

@techreport{Yungster2026,
  author  = {Yungster, S.},
  title   = {Optimization of a 10-Step Reduced {CH$_4$/O$_2$} Combustion Mechanism for {RDRE} Applications},
  institution = {NASA Glenn Research Center}, type = {NASA Contractor Report}, number = {NASA/CR-20260003291}, year = {2026}}

@misc{Cantera,
  author  = {Goodwin, D. G. and Moffat, H. K. and Schoegl, I. and Speth, R. L. and Weber, B. W.},
  title   = {Cantera: An object-oriented software toolkit for chemical kinetics, thermodynamics, and transport processes},
  year    = {2025}, note = {Version 3.2.0, https://www.cantera.org}, doi = {10.5281/zenodo.17620923}}

@article{ChemKED,
  author  = {Weber, B. W. and Niemeyer, K. E.},
  title   = {{ChemKED}: A human- and machine-readable data standard for chemical kinetics experiments},
  journal = {Int. J. Chem. Kinet.}, volume = {50}, number = {3}, pages = {135--148}, year = {2018}, doi = {10.1002/kin.21142},
  note    = {Database: https://github.com/pr-omethe-us/ChemKED-database}}

@article{HuanMarzouk2013,
  author  = {Huan, X. and Marzouk, Y. M.},
  title   = {Simulation-based optimal {Bayesian} experimental design for nonlinear systems},
  journal = {J. Comput. Phys.}, volume = {232}, pages = {288--317}, year = {2013}}

@article{Huan2024,
  author  = {Huan, X. and Jagalur, J. and Marzouk, Y.},
  title   = {Optimal experimental design: formulations and computations},
  journal = {Acta Numer.}, volume = {33}, pages = {715--840}, year = {2024}}

@article{WangSheen2015,
  author  = {Wang, H. and Sheen, D. A.},
  title   = {Combustion kinetic model uncertainty quantification, propagation and minimization},
  journal = {Prog. Energy Combust. Sci.}, volume = {47}, pages = {1--31}, year = {2015}}

@article{LivengoodWu1955,
  author  = {Livengood, J. C. and Wu, P. C.},
  title   = {Correlation of autoignition phenomena in internal combustion engines and rapid compression machines},
  journal = {Symp. (Int.) Combust.}, volume = {5}, number = {1}, pages = {347--356}, year = {1955},
  doi     = {10.1016/S0082-0784(55)80047-1}}

@inproceedings{Teasley2023,
  author  = {Teasley, T. W. and Fedotowsky, T. M. and Gradl, P. R. and Austin, B. L. and Heister, S. D.},
  title   = {Current state of {NASA} continuously rotating detonation cycle engine development},
  booktitle = {AIAA SciTech 2023 Forum}, address = {National Harbor, MD}, year = {2023}, note = {NASA NTRS 20220018157}}

@article{Ayers2023,
  author  = {Ayers, Z. M. and Athmanathan, V. and Meyer, T. R. and Paxson, D. E.},
  title   = {Variably premixed rotating detonation engine for evaluation of detonation cycle dynamics},
  journal = {J. Propuls. Power}, volume = {39}, number = {3}, year = {2023}, doi = {10.2514/1.B38868}}

@phdthesis{Huff2025,
  author  = {Huff, R. T.},
  title   = {Loss Quantification and Reduced Order Modeling of Rotating Detonation Combustor Performance},
  school  = {University of Michigan}, year = {2025}, type = {{PhD} thesis (Aerospace Engineering)}}

@article{Stechmann2019,
  author  = {Stechmann, D. P. and Sardeshmukh, S. and Heister, S. D. and Mikoshiba, K.},
  title   = {Role of ignition delay in rotating detonation engine performance and operability},
  journal = {J. Propuls. Power}, volume = {35}, number = {1}, pages = {125--140}, year = {2019}, doi = {10.2514/1.B37117}}

@inproceedings{Fievisohn2019,
  author  = {Fievisohn, R. T. and Hoke, J. L. and Holley, A.},
  title   = {Thermodynamic and operability implications of product recirculation in rotating detonation engines},
  booktitle = {AIAA Propulsion and Energy 2019 Forum}, year = {2019}, note = {AIAA 2019-4449}, doi = {10.2514/6.2019-4449}}

@inproceedings{Fievisohn2020,
  author  = {Fievisohn, R. T. and Hoke, J. and Schumaker, S. A.},
  title   = {Product recirculation and incipient autoignition in a rotating detonation engine},
  booktitle = {AIAA SciTech 2020 Forum}, publisher = {American Institute of Aeronautics and Astronautics}, year = {2020},
  doi     = {10.2514/6.2020-2286}}

@article{Nagy2015,
  author  = {Nagy, T. and Valk{\'o}, {\'E}. and Sedy{\'o}, I. and Zs{\'e}ly, I. Gy. and Pilling, M. J. and Tur{\'a}nyi, T.},
  title   = {Uncertainty of the rate parameters of several important elementary reactions of the {H$_2$} and syngas combustion systems},
  journal = {Combust. Flame}, volume = {162}, number = {5}, pages = {2059--2076}, year = {2015},
  doi     = {10.1016/j.combustflame.2015.01.005}}

@article{Vehtari2017,
  author  = {Vehtari, A. and Gelman, A. and Gabry, J.},
  title   = {Practical {Bayesian} model evaluation using leave-one-out cross-validation and {WAIC}},
  journal = {Stat. Comput.}, volume = {27}, pages = {1413--1432}, year = {2017}, doi = {10.1007/s11222-016-9696-4}}

@article{Huang2004,
  author  = {Huang, J. and Hill, P. G. and Bushe, W. K. and Munshi, S. R.},
  title   = {Shock-tube study of methane ignition under engine-relevant conditions: experiments and modeling},
  journal = {Combust. Flame}, volume = {136}, number = {1--2}, pages = {25--42}, year = {2004}, doi = {10.1016/j.combustflame.2003.09.002}}

@article{Burke2015,
  author  = {Burke, U. and Somers, K. P. and O'Toole, P. and Zinner, C. M. and Marquet, N. and Bourque, G. and Petersen, E. L. and Metcalfe, W. K. and Serinyel, Z. and Curran, H. J.},
  title   = {An ignition delay and kinetic modeling study of methane, dimethyl ether, and their mixtures at high pressures},
  journal = {Combust. Flame}, volume = {162}, number = {2}, pages = {315--330}, year = {2015}, doi = {10.1016/j.combustflame.2014.08.014}}

@article{Mathieu2015,
  author  = {Mathieu, O. and Pemelton, J. M. and Bourque, G. and Petersen, E. L.},
  title   = {Shock-induced ignition of methane sensitized by {NO$_2$} and {N$_2$O}},
  journal = {Combust. Flame}, volume = {162}, number = {8}, pages = {3053--3070}, year = {2015}, doi = {10.1016/j.combustflame.2015.03.024}}
\end{document}